\documentclass[usenatbib,usegraphicx]{mn2e}
\usepackage{amsfonts}
\usepackage{amsmath}
\usepackage{amssymb}
\usepackage{comment}
\usepackage[usenames,dvipsnames]{color}

\title[Minor-merger-driven evolution of disk galaxies]
    {The importance of minor-merger-driven star formation and black-hole growth in
    disk galaxies}
\author[Sugata Kaviraj]
{Sugata Kaviraj\thanks{s.kaviraj@herts.ac.uk}\\
Centre for Astrophysics Research, University of Hertfordshire,
College Lane, Hatfield, Herts, AL10 9AB, UK\\
Department of Physics, University of Oxford, Keble Road, Oxford,
OX1 3RH, UK}

\begin{document}

\maketitle

\def \aj {AJ}
\def \mnras {MNRAS}
\def \pasp {PASP}
\def \apj {ApJ}
\def \apjs {ApJS}
\def \apjl {ApJL}
\def \aap {A\&A}
\def \nat {Nature}
\def \araa {ARAA}
\def \iaucirc {IAUC}
\def \aaps {A\&A Suppl.}
\def \qjras {QJRAS}
\def \na {New Astronomy}
\def \aapr {A\&ARv}
\def\lesssim{\mathrel{\hbox{\rlap{\hbox{\lower4pt\hbox{$\sim$}}}\hbox{$<$}}}}
\def\gtrsim{\mathrel{\hbox{\rlap{\hbox{\lower4pt\hbox{$\sim$}}}\hbox{$>$}}}}


\begin{abstract}
We use the SDSS Stripe 82 to empirically quantify the stellar-mass
and black-hole growth triggered by minor mergers in local spiral
(disk) galaxies. Since major mergers destroy disks and create
spheroids, morphologically disturbed spirals are likely remnants
of minor mergers. Disturbed spirals exhibit enhanced specific star
formation rates (SSFRs), the enhancement increasing in galaxies of
`later' morphological type (which have more gas and smaller
bulges). By combining the SSFR enhancements with the fraction of
time spirals spend in this `enhanced' mode, we estimate that
$\sim$40\% of the star formation in local spirals is directly
triggered by minor mergers. The disturbed spirals also exhibit
higher nuclear-accretion rates, implying that minor mergers
enhance the growth rate of the central black hole. However, the
specific accretion rate shows a lower enhancement than that in the
SSFR, suggesting that the coupling between stellar-mass and
black-hole growth is weak in minor-merger-driven episodes. Given
the significant fraction of star formation that is triggered by
minor mergers, this weaker coupling may contribute to the large
intrinsic scatter observed in the stellar vs. black-hole mass
relation in spirals. Combining our results with the star formation
in early-type galaxies -- which is minor-merger-driven and
accounts for $\sim$14\% of the star formation budget -- suggests
that \emph{around half of the star formation activity in the local
Universe is triggered by the minor-merger process.}
\end{abstract}


\begin{keywords}
galaxies: formation -- galaxies: evolution -- galaxies:
interactions -- galaxies: spirals
\end{keywords}


\section{Introduction}
The standard $\Lambda$CDM model of galaxy formation postulates a
hierarchical growth of structure over cosmic time
\citep[e.g.][]{White1978,Somerville1999,Cole2000,Hatton2003}. In
this paradigm, galaxies and their host dark-matter halos grow,
both through mergers with systems of similar mass (`major
mergers'), and via the accretion of smaller objects (`minor
mergers'). A natural consequence of the shape of the observed
luminosity function -- in which smaller galaxies far outnumber
their more massive counterparts
\citep[e.g.][]{Cole2001,Blanton2001} -- is that minor mergers are
significantly more common than major interactions.

While past studies of merging have largely explored the effects of
major mergers
\citep[e.g.][]{Darg2010a,Darg2010b,Toomre1972,Hernquist1989,Barnes1992a,Jesseit2007},
a growing literature is highlighting the important role of minor
mergers (mass ratios $\lesssim$ 1:4) in influencing massive-galaxy
evolution. Minor merging is a frequent process, that is both
predicted
\citep[e.g.][]{Maller2006,Stewart2008,Fakhouri2008,Kaviraj2009}
and observed \citep[e.g.][]{Lin2004,Jogee2009,Lopez2010} to be at
least $\sim$3-4 times more common than major interactions at late
epochs. While each individual minor interaction may have a
relatively small effect on the massive galaxy, it is becoming
increasingly clear that the \emph{cumulative} impact of minor
mergers over cosmic time is significant, both in terms of the
stellar content of massive galaxies and their structural
properties.

\begin{figure*}
\begin{minipage}{172mm}
\begin{center}
\includegraphics[width=\textwidth]{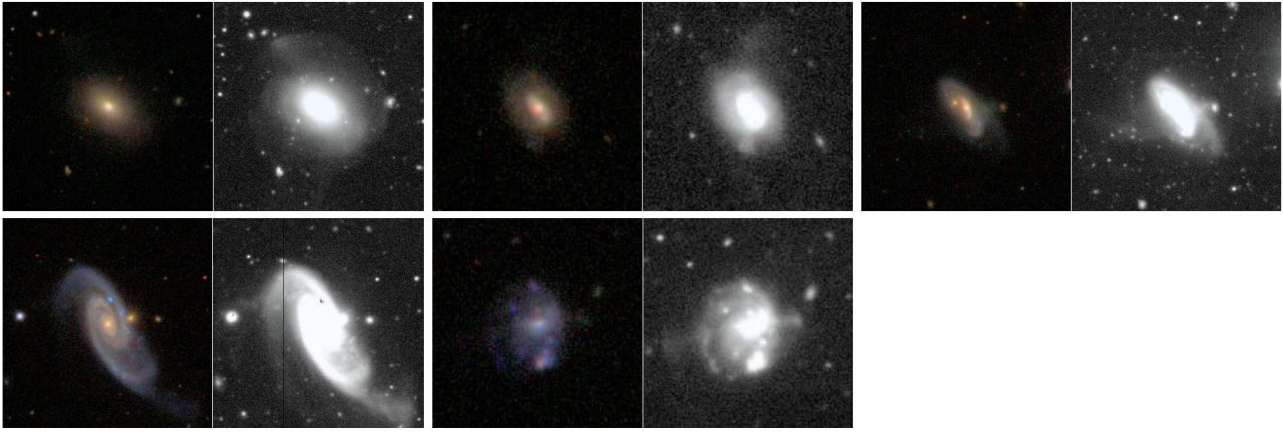}
\caption{Examples of standard-depth, multi-colour images (left)
and their deeper $r$-band Stripe 82 counterparts (right) of spiral
minor-merger remnants. We show examples of early-type (row 1,
left), Sa (row 1, middle), Sb (row 1, right), Sc (row 2, left) and
Sd (row 2, middle) galaxies. The morphological disturbances are
typically clearer in the Stripe 82 imaging, making these deeper
images necessary for this exercise. While ETGs are not considered
in this study, we show an example of a disturbed ETG for
comparison to the spirals. Note that the images may look better on
screen than in print.} \label{fig:hubble_sequence}
\end{center}
\end{minipage}
\end{figure*}

The role of this process in fuelling star formation in massive
galaxies has recently been explored via ultraviolet (UV) studies
of early-type galaxies (ETGs). Survey-scale UV data, from GALEX at
$z\sim0$ (e.g. Yi et al. 2005; Kaviraj et al. 2007) and deep
optical surveys at intermediate redshift
\cite[e.g.][]{Kaviraj2008b}, have revealed widespread star
formation in ETGs at late epochs. A strong correlation is observed
between the presence of morphological disturbances and blue UV
colours in nearby ETGs (Kaviraj et al. 2011; see also Schweizer et
al. 1990, 1992), indicating that the star formation is driven by
mergers, and not by processes like stellar mass loss or accretion
that would leave the stellar morphology of the system undisturbed
(Kaviraj et al. 2011, see also Crockett et al. 2011). However, the
major-merger rate at late epochs ($z<1$) cannot satisfy the number
of disturbed ETGs, suggesting that \emph{minor} mergers dominate
the star formation in these objects (Kaviraj et al. 2011, see also
Shabala et al. 2012). Note that these studies primarily sample
low-density environments, making these results relevant largely to
ETGs that inhabit groups and the field (and not dense clusters).
However, since clusters are typically hostile to star formation
\citep[e.g.][]{Dressler1984,Moore1999,Kimm2011}, the bulk of the
ETG star formation budget is expected to reside in low-density
environments.

While past empirical work on minor-merger-driven star formation
has focused on ETGs, this process will clearly affect massive
galaxies regardless of morphology. \citet[][K14
hereafter]{Kaviraj2014} has leveraged the ETG studies mentioned
above to derive a lower limit for the fraction of star formation
in spiral galaxies that is also likely triggered by minor mergers.
The K14 results indicate that \emph{at least} a quarter of the
star formation budget in local spirals is attributable to this
process, yielding a lower limit for the minor-merger-driven
fraction of \emph{cosmic} star formation of $\sim$35\%. The
observed correlation between black hole and galaxy stellar mass
\citep[e.g.][]{Gultekin2009} further implies that a similar
fraction of black-hole growth may also be induced by this process.

The work on minor-merger-driven star formation is mirrored by
studies of the size evolution of massive galaxies over cosmic
time. Recent empirical work has demonstrated that the effective
radii of massive galaxies increase, on average, by factors of 3-5
over the lifetime of the Universe
\citep[e.g.][]{Daddi2005,Trujillo2006,VD2008,Buitrago2008,Saracco2009,Cimatti2012,Newman2012,Ryan2012,Huertas-Company2012}.
While secular mechanisms, such as adiabatic expansion driven by
stellar mass loss or strong AGN feedback have been proposed to
explain this evolution \citep[e.g.][but see Trujillo et al. 2011
and Bluck et al. 2012]{Fan2008,Damjanov2009}, consensus favours
size evolution via mergers, with the bulk of the increase
attributed to minor mergers over cosmic time
\citep[e.g.][]{Khochfar2006,Bournaud2007,Naab2009,Nipoti2009,Oser2012}.

While it is increasingly well-established that the minor-merger
process can significantly influence the star-formation and
structural properties of massive galaxies, its overall impact on
cosmic star formation and black hole growth remains poorly
understood. Since spirals dominate the star-formation budget at
late epochs (K14), this translates into a need to understand, in
detail, the impact of minor mergers on the spiral population
across cosmic time. While K14 has derived a lower limit
($\sim$25\%) for the fraction of star formation in spirals that is
triggered by this process at the present day, an estimate of its
global impact on galaxy evolution demands a more accurate
measurement of this fraction, beyond the lower limit derived by
K14.

A burgeoning literature has now begun probing the impact of minor
mergers on spiral galaxies. Empirical studies of spiral
minor-merger \emph{remnants} have largely focused on individual
case studies of very nearby galaxies. These efforts typically
demonstrate enhancements in star formation and, in some cases,
nuclear activity \citep[see
e.g.][]{Smith1996,Knapen2004,Mazzuca2006}, indicating that the
minor-merger process is indeed influencing the growth of both the
galaxy and its central black hole. {\color{black}Studies of minor
galaxy \emph{pairs} -- i.e. systems where the two objects have not
coalesced -- also show enhancements, especially when the system is
close to coalescence (e.g. separations $<10$ kpc), although the
effect is much weaker than in major mergers
\citep{Li2008,Ellison2008,Robaina2009,Scudder2012,Patton2013}.}
The triggering of star formation in the massive progenitor of a
minor merger is expected to peak at the `remnant' stage, during
the coalescence of the satellite, when the tidal forces on the
disk are highest \citep[see e.g.][]{Mihos1994,Kaviraj2009} -
statistical studies of minor-merger \emph{remnants} are,
therefore, very desirable for gaining insights into this process.

\begin{figure*}
\begin{minipage}{172mm}
\begin{center}
\includegraphics[width=\textwidth]{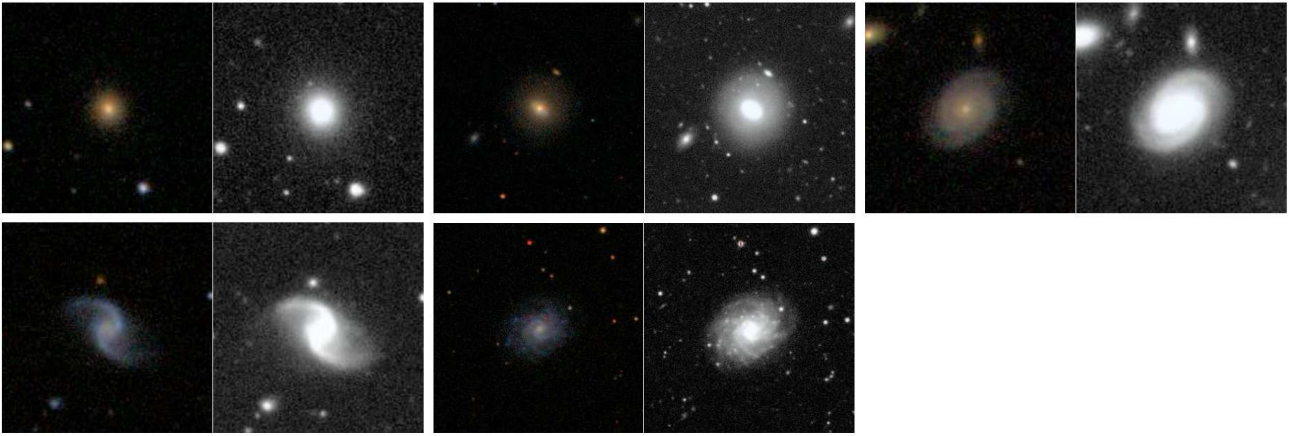}
\caption{Examples of standard-depth, multi-colour (left) images
and their deeper $r$-band Stripe 82 counterparts (right) of
relaxed spiral galaxies. Note the lack of tidal disturbances
compared to the objects shown in Figure 1 above. We show examples
of early-type (row 1, left), Sa (row 1, middle), Sb (row 1,
right), Sc (row 2, left) and Sd (row 2, middle) galaxies. While
ETGs are not considered in this study, we show an example of an
ETG for comparison to the spirals. Note that the images may look
better on screen than in print.} \label{fig:minor_merger_remnants}
\end{center}
\end{minipage}
\end{figure*}

In a similar vein, while past theoretical efforts have largely
explored the impact of \emph{major} mergers, recent work has begun
to study the role of minor interactions in spiral galaxies. In
qualitative agreement with the empirical case studies mentioned
above, theoretical efforts do find star-formation enhancement in
minor-merger remnants. The enhancement typically anti-correlates
with the system mass ratio
\citep[e.g.][]{Hernquist1989,Mihos1994,Hernquist1995,Cox2008}, but
the actual increase in star-formation efficiency depends
critically on the specific theoretical prescriptions utilized in
individual simulations.

Since the enhancement of star formation occurs at the `remnant'
stage of the process i.e. when the smaller companion is coalescing
with the primary, an \emph{empirical} survey-scale study of spiral
minor-merger remnants becomes a compelling exercise. In this paper
we present such a study, whose principal aims are to (1) quantify
the enhancement of star formation due to minor mergers in spiral
galaxies (2) estimate the fraction of total star formation in
spirals, and in massive galaxies in general, that is triggered by
minor mergers and (3) quantify the enhancement of black hole
growth in spirals due to the minor-merger process. This study
significantly extends the K14 analysis (which derived a robust
lower limit for point [2] above), with the overall objective of
gauging the global impact of minor mergers on the evolution of the
spiral galaxy population.

This paper is organized as follows. In Section 2, we describe the
galaxy sample that underpins this study and the selection of
minor-merger remnants via visual inspection. In Sections 3 and 4,
we explore the enhancement of star formation in spirals due to
minor mergers and use these results to estimate the fractions of
the spiral and cosmic star-formation budgets that are triggered by
the minor-merger process. In Section 5, we explore
minor-merger-driven enhancement of nuclear accretion in spiral
galaxies. We summarize our findings in Section 6. Throughout this
study we use the \emph{WMAP3} cosmological parameters
\citep{Komatsu2011}: $\Omega_m = 0.241$, $\Omega_{\Lambda} =
0.759$, $h=0.732$, $\sigma_8 = 0.761$.


\section{Data}
\subsection{Galaxy sample}
The galaxies used in this study are based on a sample that was
recently compiled by K14. The objects are drawn from the Sloan
Digital Sky Survey (SDSS) Stripe 82, a $\sim300$ deg$^2$ region
along the celestial equator in the Southern Galactic Cap
$(-50^{\circ} < \alpha < 59^{\circ}, -1.25^{\circ} < \delta <
1.25^{\circ})$ \citep{Frieman2008} that offers a co-addition of
122 imaging runs \citep{York2000,Abazajian2009}, yielding images
that are $\sim$2 mags deeper than the standard-depth, 54 second
SDSS scans (which have magnitude limits of $22.2$, $22.2$ and
$21.3$ mags in the $g$, $r$ and $i$-bands respectively).

K14 classified their galaxies into standard morphological classes
\citep[E/S0, Sa, Sb, Sc, Sd, see][]{Hubble1926,dev1959}, using
visual inspection of both the standard-depth, colour images from
the SDSS DR7 and their deeper $r$-band Stripe 82 counterparts.
While morphological parameters, such as Concentration, Asymmetry,
Clumpiness, M$_{20}$ and the Gini coefficient, have often been
employed for morphological classification of survey datasets
\citep[e.g.][]{Conselice2003,Lotz2004}, they are calibrated
against results from visual inspection
\citep[e.g.][]{Abraham1996}, which yields better precision in the
classification of galaxy morphologies. Leveraging the findings of
recent work that has employed visual inspection of SDSS galaxies
\citep[see
e.g.][]{Kaviraj2007c,Schawinski2007,Fukugita2007,Lintott2008,Nair2010,Lintott2011},
K14 restricted their sample to $r<16.8$ and $z<0.07$, where
morphological classification from SDSS images is likely to be most
reliable. The final K14 compilation contains $\sim$6,500 galaxies
in this redshift and magnitude range.

In our analysis below, we employ published stellar masses
\citep{Kauffmann2003} and star formation rates
\citep[SFRs;][]{Brinchmann2004} from the latest version of the
publicly-available MPA-JHU value-added SDSS
catalogue\footnote{http://www.mpa-garching.mpg.de/SDSS/DR7/}.
Briefly, stellar masses are calculated by comparing $ugriz$
photometry of individual galaxies to a large grid of synthetic
star formation histories, based on the \cite{BC2003} stellar
models. Model likelihoods are calculated from the values of
$\chi^2$, and 1D probability distributions for free parameters,
such as stellar masses, are constructed via marginalization. The
median of this 1D distribution is taken to be the best estimate
for the parameter in question, with the the 16th and 84th
percentile values (which enclose 68\% of the total probability)
yielding a `1$\sigma$' uncertainty.

The SFRs are estimated via two different methods, depending on the
ionization class of the galaxy, that is derived using an optical
emission-line ratio analysis \citep[][see also Baldwin et al.
1981, Veilleux et al. 1987, Kauffmann et al. 2003]{Kewley2006},
using the values of [NII]/H$\alpha$ and [OIII]/H$\beta$ measured
from the SDSS spectra of individual galaxies
\citep{Brinchmann2004,Tremonti2004}. Objects in which all four
emission lines are detected with a signal-to-noise ratio greater
than 3 are classified as either `star-forming', `composite' (i.e.
hosting both star formation and AGN activity), `Seyfert' or
`LINER', depending on their location in the [NII]/H$\alpha$ vs.
[OIII]/H$\beta$ diagram \citep[see e.g.][]{Kewley2006}. Galaxies
without a detection in all four lines are classified as
`quiescent'.

SFRs for galaxies classified as `star-forming' (i.e. where the
nuclear ionization is driven by star formation) are estimated by
comparing galaxy spectra to a library of models from
\citet{Charlot2001}, with a dust treatment that follows the
empirical model of \citet{Charlot2000}. For galaxies that are
classified as `quiescent' , or those that are classified as
`AGN/Composite' (in which a significant fraction of the nuclear
emission is likely driven by a central AGN), SFRs are calculated
using the D4000 break\footnote{The D4000 break is produced by the
absorption of short-wavelength photons from metals in stellar
atmospheres. The feature is stronger in systems that are deficient
in hot, blue stars \citep[e.g.][]{Poggianti1997}, making it a good
estimator of SFR in systems with weak emission lines, or those in
which the relevant diagnostic lines, such as H$\alpha$, are
contaminated by non-thermal sources (e.g. AGN)}. The correlation
between specific SFR and D4000 break for the `star-forming'
subsample is used to estimate the specific SFR of the galaxy in
question. Note that the SFRs used here are corrected for both
internal extinction and the fixed size of the SDSS fibre. We refer
readers to \citet{Brinchmann2004} for full details of the
modelling.


\subsection{Selection of minor-merger remnants}
Theoretical work indicates that the strong gravitational torques
induced by major mergers (mass ratios $\gtrsim1:4$) destroy
rotationally-supported disks and produce pressure-supported
spheroidal systems
\citep[e.g.][]{Toomre1972,Toomre1977,Hernquist1989,Barnes1992a,Cox2006,Naab2006,Jesseit2007,Hopkins2008}.
While major interactions involving very high gas fractions may
yield late-type remnants that rebuild disks from the remnant gas
\citep[e.g.][]{Springel2005a,Robertson2006,Hopkins2009}, such
conditions are rare at low redshift \citep{Kannappan2004}. Thus,
systems in the local Universe that exhibit both disk morphology
and tidal disturbances are likely to be remnants of minor mergers,
since a major interaction would have destroyed the disk and
created a spheroid.

While disturbed spirals are likely minor-merger remnants, visual
identification of such systems requires deep, high-resolution
colour imaging. Deep images are required for detecting the tidal
features that result from recent minor interactions
\citep[e.g.][]{Kaviraj2010}, while high-resolution, colour images
enable us to efficiently differentiate disks, which have
inhomogeneous structure, from the smooth, homogeneous light
distributions of early-type (i.e. spheroidal) galaxies. The
combination of standard-depth colour images and co-added scans
makes the SDSS Stripe 82 an ideal platform for a large-scale study
of spiral minor-merger remnants. To identify our sample, we
visually inspect both the standard-depth colour image and the
$r$-band Stripe 82 counterpart of each K14 galaxy and flag those
that are morphologically disturbed.

Figure 1 shows examples of disturbed spirals and, for comparison,
their counterparts in the ETG population. The morphological
disturbances are most easily detected in the Stripe 82 imaging,
making these deep images indispensable for this exercise. Figure 2
shows examples of relaxed spirals and ETGs, that do not show
evidence for disturbed morphologies in the SDSS images. Note that,
while previous work has flagged morphological disturbances via
visual inspection \citep[e.g.][]{Nair2010}, these have been based
on standard-depth SDSS images. Since the faint tidal debris
produced by minor mergers \citep[e.g.][]{Peirani2010} is only
visible in deeper imaging, such as that available in the Stripe 82
\citep{Kaviraj2010}, using just the standard-depth images would
significantly underestimate the fraction of galaxies that have
undergone recent minor interactions.


\section{Minor-merger-driven star formation in spiral galaxies}

\subsection{Enhancement of star formation due to minor mergers}
We begin by setting up a theoretical framework to quantify the
fraction of star formation in spiral galaxies that is triggered by
the minor-merger process.

\begin{table*}
\begin{center}
\caption{Columns from left to right: (1) Morphological class (2)
duty cycle ($D$ in Eqn 1) of the remnant phase of the minor-merger
process i.e. the fraction of time the galaxy appears disturbed due
to minor mergers (3) enhancement of specific star formation rate
in remnant phase of the minor-merger process ($\eta$ in Eqn 1) (4)
the fraction of star formation driven by minor mergers, calculated
using Eqn. 1 (5) proportion of the star formation budget in spiral
galaxies that is hosted by this morphological class (from K14).
{\color{black}Note that the star formation budgets calculated by
K14 include corrections for Malmquist bias, via the standard
1/V$_{max}$ weighting method.}}
\begin{tabular}{lllll}

\textbf{Morphology} & \textbf{$D$} & \textbf{$\eta$} &
\textbf{$F_{mm}$ (via Eqn. 2)} & \textbf{Proportion of spiral SF
budget (from K14)}\\\hline
Sa                  & $0.16$       & 1.98            & 0.27              & 0.19\\
Sb                  & $0.17$       & 3.62            & 0.43              & 0.34\\
Sc                  & $0.13$       & 6.15            & 0.48              & 0.32\\
Sd/Irr              & $0.11$       & 6.14            & 0.43              & 0.15\\
\end{tabular}
\end{center}
\end{table*}

We define the following quantities: $\phi_0$ is the normal
specific star-formation rate of a spiral galaxy in the absence of
a minor merger, $m$ is its stellar mass, $D$ is the duty cycle of
the remnant phase of the minor-merger process i.e. the fraction of
time the galaxy appears disturbed due to minor mergers and $\eta$
is the enhancement in the specific star formation rate while the
galaxy is in this disturbed phase. The stellar mass ($S$) produced
in a time period ($\delta t$) can then be expressed as:

\begin{equation}
S = \underbrace{\phi_0.(1-D).m.\delta
t}_\text{S$_{\textnormal{NORM}}$} \hspace{0.05in} +
\hspace{0.05in} \underbrace{\eta.\phi_0.D.m.\delta
t}_\text{S$_{\textnormal{MM}}$}
\end{equation}

S$_{\textnormal{NORM}}$ represents the stellar mass formed while
the galaxy is in a normal star-forming mode (i.e. in the absence
of a minor merger), while S$_{\textnormal{MM}}$ represents the
stellar mass formed while the galaxy experiences enhanced star
formation during the remnant phase of a minor merger. The fraction
of star formation triggered by minor mergers ($F_{MM}$) is then:

\begin{equation}
F_{\textnormal{MM}} = \frac{S_{\textnormal{MM}}}{S} =
\frac{\eta.D}{1+D.(\eta-1)}
\end{equation}

As might be expected, $F_{MM}$ depends on a combination of the
enhancement of star formation during the disturbed phase and the
fraction of time that the galaxy spends in this phase. While we
cannot measure these quantities for individual galaxies, we can
calculate \emph{mean} statistical estimates for $\eta$ and $D$, if
large populations of minor-merger remnants are available, as is
the case in this study. As K14 has suggested, the star formation
enhancement is likely to be higher in `later' morphological types,
which host larger gas reservoirs \citep[e.g.][]{Kannappan2004} and
smaller bulges that are less able to stabilize the disk against
gas inflows \citep[e.g.][]{Mihos1994,Hernquist1995,Martig2009}. It
is, therefore, desirable to perform this analysis as a function of
morphological type.

Assuming that the detectability of morphological disturbances does
not evolve in the short redshift range studied here, the duty
cycle ($D$) is given by the fraction of galaxies that are
disturbed. In column 2 of Table 1, we present the fraction of
disturbed spirals as a function of morphology. We find similar
duty cycles ($\sim$11-17\%) across different morphological types.
The values for Sa and Sb systems is similar to that found in the
massive ETG population \citep[18\%; see][]{Kaviraj2010}. The
decreasing duty cycles towards `later' morphological types is
likely driven by the average galaxy mass being lower in these
classes, which affects the ability of the galaxy to attract
satellites.

\begin{figure}
\begin{center}
$\begin{array}{c}
\includegraphics[width=2.95in]{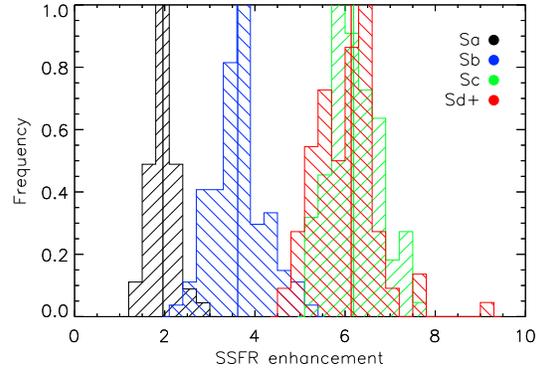}
\end{array}$
\caption{The SSFR enhancement in disturbed spirals in different
morphological classes: S0 (far left), Sa (second from left), Sc
(histogram on the right, with hatching from bottom left to top
right), Sd and later (histogram on the right, with hatching from
bottom right to top left). The enhancement is defined as the ratio
of the median SSFR of the disturbed objects to that in a control
(undisturbed) sample that has the same redshift and $r$-band
distribution. See text in Section 3.1 for details.}
\label{fig:u_r}
\end{center}
\end{figure}

We calculate mean SSFR enhancements ($\eta$) using the published
MPA-Garching SSFRs of the relaxed and disturbed spirals in each
morphological class. To perform this comparison statistically, we
extract, in each morphological class (Sa/Sb/Sc etc.), 1000
{\color{black}random} samples of relaxed galaxies, each with the
same redshift and $r$-band distribution as that of the disturbed
systems {\color{black}(the samples are not necessarily
independent)}. Matching in redshift and magnitude implies that we
are comparing galaxies at similar distances and with similar
luminosities, making the comparison more meaningful. We then
calculate the median SSFR of each `matched' relaxed sample and
construct the ratio (disturbed:relaxed) of median SSFRs for each
comparison, which we define as the SSFR enhancement. This yields,
for each morphological class, a distribution of
{\color{black}1000} SSFR enhancements (Figure 3). We take the
median of this distribution as a typical value for $\eta$ for the
morphological class in question.

Figure 3 indicates that the SSFR is enhanced in minor-merger
remnants regardless of galaxy morphology, the enhancement
increasing in galaxies that have `later' morphological type. While
disturbed Sa galaxies exhibit median enhancements of around a
factor of $\sim$2, this climbs to a factor of $\sim$6 for Sc
galaxies and later morphological classes. The values of $\eta$ are
presented in column 3 of Table 1.


\subsection{The fraction of disk and cosmic star formation that is triggered
by minor mergers} Given the values of $\eta$ and $D$ calculated
above, we use Eqn. 2 to estimate the fraction of star formation
that is triggered by minor mergers ($F_{MM}$) in each
morphological class (column 4 in Table 1). The values of $F_{MM}$
are $\sim$0.27 for Sa galaxies and climb to $\sim$0.43-0.48 in Sb
and later morphological types. Multiplying the values of $F_{MM}$
with the proportion of the star-formation budget in spirals that
is hosted by each morphological class (column 5 in Table 1, taken
from K14), then yields an estimate for the \emph{total} fraction
of star formation in spiral galaxies that is attributable to minor
mergers. We estimate this fraction to be $\sim$40\% (consistent
with the results of K14, who estimated an empirical lower limit
for this value of $\sim$25\%.) \emph{Thus, a significant fraction
of stellar mass in local spirals is likely to be created in
enhanced star-formation episodes during the latter stages of minor
mergers.}

We note that our analysis has only considered systems that are
visibly morphologically disturbed, and not those that may have
experienced a recent minor merger, but where the mass ratio is not
sufficiently high to induce morphological disturbances at the
depth of the SDSS images. While we assume that the star formation
enhancement in such mergers is negligible, the robustness of this
assumption is difficult to test without dedicated theoretical
simulations, which are beyond the scope of this paper.

{\color{black}In addition, given the evidence for star formation
enhancement in minor galaxy \emph{pairs} described above, an
(additional) contribution to minor-merger-driven star formation is
expected while the galaxies are at small separations but still `on
approach'. However, given a duty cycle of $\sim$2\% and typical
enhancement of a factor of $\sim$3 \citep{Scudder2012}, the extra
contribution of this phase is around a percent and therefore does
not alter our overall conclusions.} Nevertheless, as a result of
both these points, the value of 40\% derived above may indeed also
be a lower limit.

Finally, it is worth combining our results with the star formation
in ETGs, which is driven by minor mergers (at least in low-density
environments) and accounts for $\sim$14\% of the cosmic star
formation budget (K14). This indicates that \emph{around half
(0.4$\times$0.86\% [LTGs] + 14\% [ETGs]) of the cosmic star
formation activity at the present day is triggered by the
minor-merger process.}


\section{Minor-merger-triggered nuclear accretion in spirals}
The observed correlation between galaxy stellar mass and the mass
of the central black hole
\citep[e.g.][]{Gultekin2009,McConnell2011} suggests that the black
hole and its host galaxy co-evolve \citep[e.g.][]{DiMatteo2008b}.
While the correlation is strongest in early-type systems
\citep[e.g.][]{Magorrian1998,Gebhardt2000,Ferrarese2000,Rix2004},
it also exists in spirals, albeit with larger intrinsic scatter
\citep{Gultekin2009}. Thus, for every unit of stellar mass that is
built, a certain fraction is, \emph{on average}, accreted on to
the black hole (in early-type systems this fraction is
$\sim$1/1000, see e.g. Merritt \& Ferrarese 2001).

\begin{table*}
\begin{center}
\caption{Number fractions in different emission-line classes for
relaxed and disturbed objects, split by early and late spirals
(see Section 4). See Section 2.1 for a description of how galaxies
are classified into these ionization classes (QS = Quiescent, SF =
Star forming, Low S/N SF = Low signal-to-noise star-forming, CP =
Composite, Sy = Seyfert, LI = LINER). The first three rows
correspond to the early spiral population, while the next three
rows correspond to the late spiral population. f(Dst)/f(Rel)
indicates the ratio (disturbed:relaxed) for each ionization
class.}
\begin{tabular}{lllllll}
\textbf{}               & \textbf{QS} & \textbf{SF} & \textbf{Low
S/N SF} & \textbf{CP} & \textbf{Sy} & \textbf{LI}\\\hline
Relaxed early spirals   & 0.10        & 0.39        & 0.20                & 0.12        & 0.08        & 0.11\\
Disturbed early spirals & 0.07        & 0.47        & 0.13                & 0.15        & 0.10        & 0.08\\
f(Dst)/f(Rel)           & 0.70        & 1.21        & 0.65                & 1.25        & 1.25        & 0.73\\\\

Relaxed late spirals    & 0.04        & 0.57        & 0.22                & 0.07        & 0.04        & 0.06\\
Disturbed late spirals  & 0.04        & 0.62        & 0.13                & 0.10        & 0.06        & 0.05\\
f(Dst)/f(Rel)           & 1.00        & 1.09        & 0.59                & 1.43        & 1.50        & 0.83\\
\end{tabular}
\end{center}
\end{table*}

The positive correlation between stellar and black hole mass in
spirals suggests that the enhanced star formation in the disturbed
spirals should be accompanied by enhanced nuclear accretion. It is
worth noting, however, that the larger intrinsic scatter in the
relation for spirals indicates a higher level of stochasticity in
the feeding of their central black holes, so that nuclear
accretion may not be as tightly coupled to star formation as in
ETGs. Nevertheless, it is interesting to explore whether nuclear
activity is enhanced in the disturbed spirals, implying enhanced
growth of their central black holes.

Recent work suggests that, at least in major mergers, AGN activity
(and therefore the nuclear accretion rate) peaks in the
post-merger i.e. remnant phase \citep{Carpineti2012}. It seems
reasonable to assume that the same holds true for minor mergers,
since star formation is preferentially fuelled in the remnant
phase. Hence, the enhancement of nuclear accretion during the
minor-merger process is also best explored in the remnant phase
i.e. via the disturbed spirals, as has been done for the
star-formation analysis above.

Following the recent literature
\citep[e.g.][]{Kewley2006,Schawinski2007}, we estimate the nuclear
accretion rate by exploiting the luminosity of the forbidden
\texttt{O[III]} line at 5007 \AA. This exercise is complicated by
the fact that this line has contributions from both the central
AGN and star formation. We therefore use the emission-line
analysis described in Section 2.1 to restrict our study to
Seyferts, where the ionization is dominated by the central AGN. We
do not consider LINERs, as the ionization in these systems may
have several sources, including shocks or evolved stellar
populations (such as P-AGB stars) associated with recent star
formation \citep[e.g.][]{Ho1993,Sarzi2010}. Following
\citet{Kewley2006}, we calculate a quantity that is proportional
to the specific accretion rate, by dividing the \texttt{O[III]}
luminosity by $\sigma^4$, where $\sigma$ is the stellar velocity
dispersion. While the \texttt{O[III]} luminosity scales with the
AGN's bolometric luminosity \citep{Heckman2004}, $\sigma^4$
provides an estimate of the mass of the central black hole
\citep[e.g.][]{Ferrarese2000,Gebhardt2000}.

\begin{figure}
\begin{center}
$\begin{array}{c}
\includegraphics[width=2.95in]{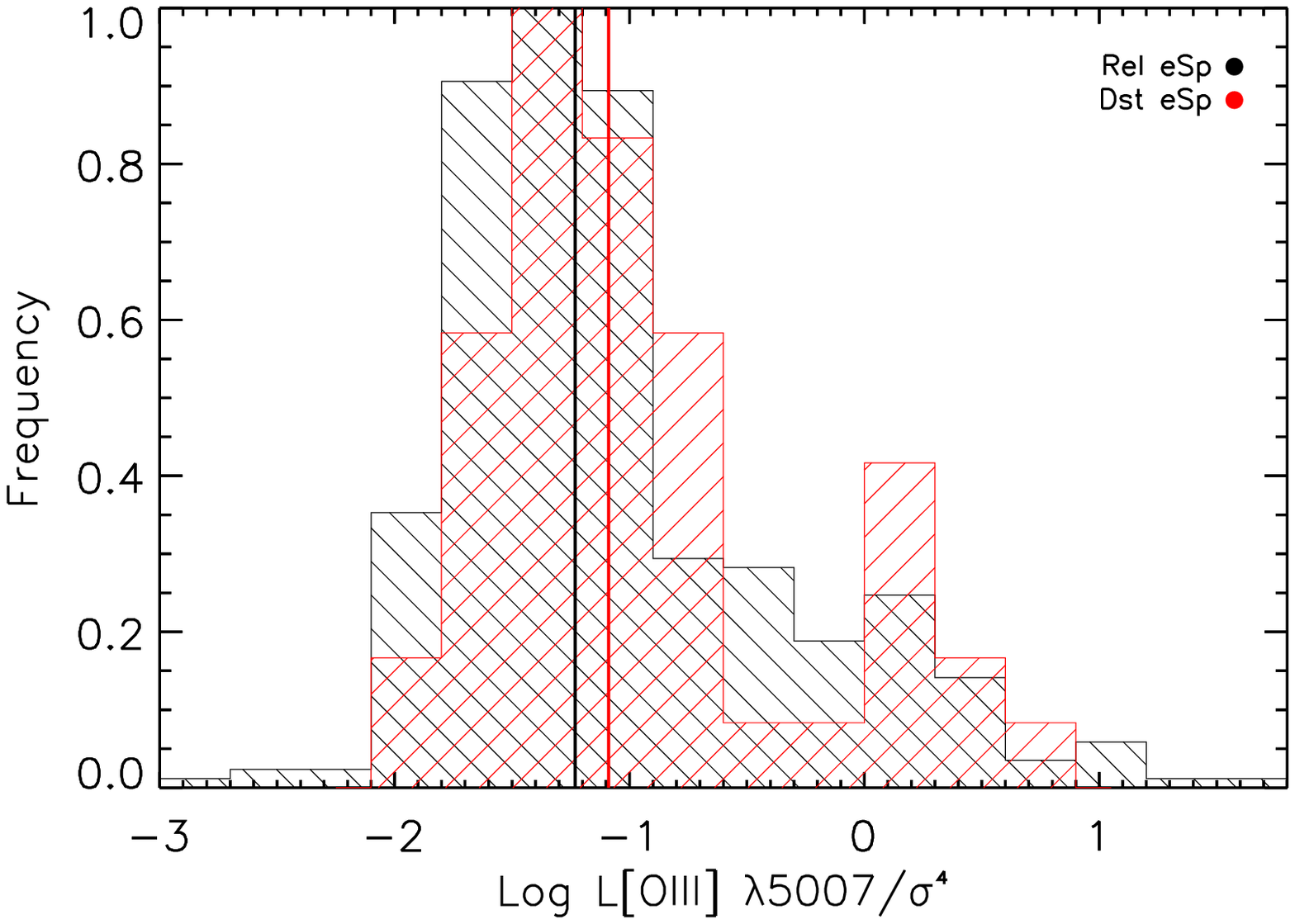}\\
\includegraphics[width=2.95in]{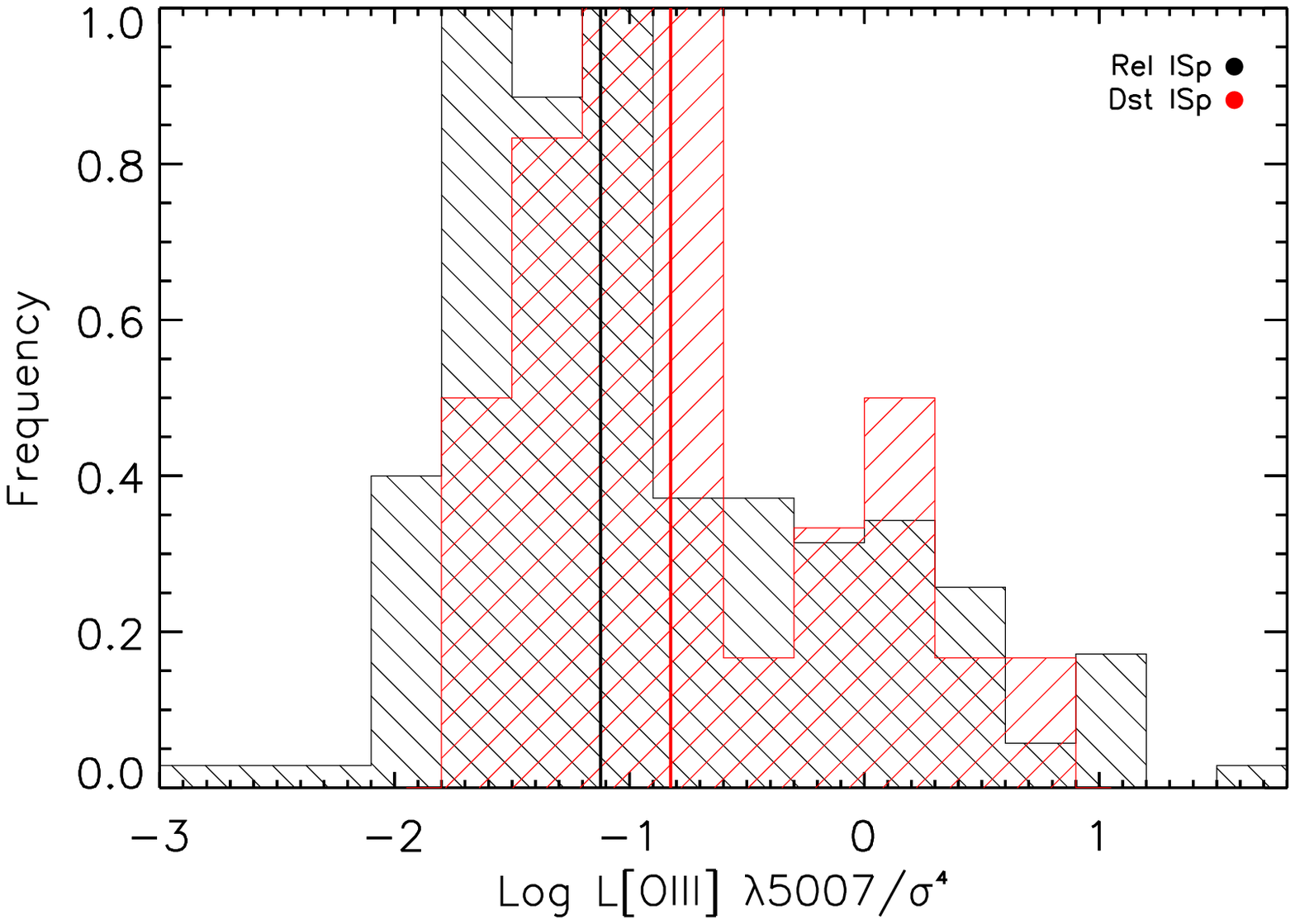}\\
\end{array}$
\caption{Enhancement in the specific nuclear accretion rate for
early (Sa) spirals (top) and late (Sb and later) spirals (bottom).
Note the log scale on the horizontal axis. The median enhancement
is relatively small ($\sim$40\%) in the early spirals, rising to a
factor of $\sim$2 in the late spirals.}\label{fig:agnenh}
\end{center}
\end{figure}

Restricting our study to Seyferts significantly reduces the number
of objects in the analysis. As a result, it is not possible to
explore the specific accretion rate as a function of fine
morphological classes (e.g. Sa/Sb/Sc etc.) as was the case in the
star-formation analysis above. Instead, we split the spiral
galaxies into early spirals (Sa systems) and late spirals (Sb and
`later' types). Table 2 indicates how the relaxed and disturbed
spirals split according to these ionization classes. The disturbed
spirals typically exhibit higher fractions of objects classified
as `star-forming', `composite' and `Seyfert', regardless of
morphology.

In Figure \ref{fig:agnenh}, we present the distribution of
specific accretion rates for these two populations (which lie in a
similar range of values to that found in recent studies of massive
galaxies e.g. Schawinski et al. 2007). We find that nuclear
accretion is indeed enhanced in the minor-merger remnants,
indicating that the minor-merger process also enhances black-hole
growth {\color{black}(KS tests on the accretion-rate distributions
of relaxed and disturbed galaxies result in probabilities of 92\%
and 98\% that they are different in the early spirals and late
spirals respectively)}. Mirroring the results of the
star-formation analysis, the enhancement is larger in the later
morphological types. While the median enhancement is relatively
mild ($\sim$40\%) in early spirals, it rises to a factor of
$\sim$2 in the late spirals.

It is worth noting, however, that the median enhancement in
nuclear accretion appears significantly less pronounced than that
in star formation activity, regardless of morphological class.
Unless nuclear accretion peaks at a different point during the
minor-merger process to star formation (which seems unlikely given
the arguments above), this suggests a weaker coupling between the
minor-merger-driven growth of stellar and black hole mass in
spiral galaxies. Given that a significant fraction of stellar mass
is likely triggered by the minor-merger process, this weaker
coupling between star formation and nuclear accretion may be
partly responsible for the larger intrinsic scatter in the stellar
vs. black-hole mass correlation in spirals.


\section{Summary}
We have used a sample of bright ($r<16.8$), local ($z<0.07$)
galaxies from the SDSS Stripe 82 to probe the role of minor
mergers in driving stellar mass and black hole growth in nearby
massive spiral (disk) galaxies. Our study has been based on a
sample of galaxies from K14, that have been visually classified
into standard morphological classes (E/S0, Sa, Sb, Sc, Sd, etc.)
using both colour images from the SDSS DR7 and their deeper
$r$-band counterparts from the Stripe 82.

Since `major' (i.e. equal or nearly equal-mass) mergers produce
spheroids, an effective route to studying spiral galaxies that are
minor-merger remnants is to use spirals that are morphologically
disturbed. The disturbed morphology indicates a recent
interaction, and the continued presence of a disk indicates that
the interaction was not a major merger (since this would have
destroyed the disk and created a spheroid). Using the DR7 and
Stripe 82 images, we have flagged spiral galaxies in the K14
sample that are morphologically disturbed, thus selecting the
nearby spiral minor-merger remnant population in the SDSS Stripe
82 field.

We have used this sample to quantify the stellar mass growth in
spiral galaxies that is plausibly triggered by minor mergers. As
indicated by Eqn. 2 and described in Section 3.1, the proportion
of minor-merger-driven star formation depends both on the
enhancement of star formation during the remnant phase of a minor
merger ($\eta$) and the fraction of time galaxies spend in this
enhanced star-formation mode (the `duty cycle', $D$). While we
cannot measure these quantities for individual galaxies, we can
calculate mean statistical estimates for $\eta$ and $D$, if large
populations of minor-merger remnants are available, as is the case
in this study.

Assuming that the detectability of morphological disturbances does
not evolve in the short redshift range studied here, $D$ can be
estimated by the fraction of galaxies that are morphologically
disturbed. $\eta$ can be calculated using the ratio of the
measured SSFRs in the relaxed spirals to that in the disturbed
spirals.

The duty cycle is relatively insensitive to morphological class
and ranges between 11 and 17\%. The SSFRs are enhanced in the
disturbed spirals, regardless of morphology, with the enhancement
increasing in galaxies that have `later' morphological type. While
disturbed Sa galaxies exhibit SSFR enhancements of a factor of
$\sim$2, this increases to a factor of $\sim$6 for Sc and later
morphological types. Our results indicate that the star-formation
enhancements are higher in galaxies that host larger internal gas
reservoirs and smaller bulges (that are less able to stabilize the
disk against radial gas inflows), as might be expected from the
recent literature.

The duty cycles and SSFR enhancements imply that the fraction of
star formation driven by minor mergers range from $\sim$27\% in Sa
galaxies to $\sim$43-48\% in Sb and later morphological types.
Combining this with the proportion of the star formation budget
hosted by each morphological class then yields \emph{a total
minor-merger-driven fraction of star formation in spirals of
$\sim$40\%}. This is consistent with the results of K14, who
estimated an empirical lower limit for this value of $\sim$25\%.
Thus, while most of the star formation in today's spiral galaxies
is unrelated to mergers, a significant fraction is attributable to
the minor-merger process. It is worth noting that our analysis has
been restricted to galaxies that show morphological disturbances.
Thus, we have disregarded systems that might have undergone a
recent minor merger, but where the mass ratio is not high to
induce morphological disturbances at the depth of the Stripe 82
imaging. {\color{black}We have also disregarded the contribution
of the phase when minor galaxy pairs are close to coalescence but
still on approach (although we have shown that this contribution
is relatively insignificant)}. Nevertheless, both these points
imply that the estimated value of 40\% above may also be a lower
limit. Combining our results with the star formation in early-type
galaxies, which is dominated by minor mergers (at least in
low-density environments) and accounts for 14\% of cosmic star
formation budget (K14), indicates that \emph{around half the star
formation activity at the present day is likely triggered by the
minor-merger process.}

The observed correlation between galaxy stellar mass and the mass
of the central black hole suggests that the enhanced star
formation episodes triggered by minor mergers may also be
accompanied by enhanced nuclear accretion. We have studied the
specific accretion rate in `early' (Sa) and `late' (Sb and later
types) spirals, by using the quantity L[OIII]/$\sigma^4$,
restricted to galaxies that are classified as Seyfert (in which
the ionization is dominated by the central AGN). Mirroring the
star formation results, the specific accretion rate is found to be
enhanced in the disturbed spirals, with the enhancement higher in
`later' morphological types. However, the enhancement in nuclear
accretion is not as pronounced as that found in the star formation
activity, suggesting that star formation and nuclear accretion may
not be tightly coupled in these minor-merger-enhanced episodes.
This relatively weak correlation may be partly responsible for the
larger intrinsic scatter in the stellar vs. black-hole mass
correlation in spiral galaxies compared to their early-type
counterparts.

Combined with the recent study of K14, the results of this paper
further highlight the important (and still poorly quantified) role
of minor mergers in influencing stellar mass and black hole growth
in massive galaxies. While a significant fraction of star
formation in \emph{nearby} spirals is triggered by minor mergers,
the role of this process is likely to become more significant with
increasing redshift, both due to the gradually increasing merger
rate \citep[e.g.][]{Lotz2011} and the higher gas fractions in
massive galaxies at earlier epochs \cite[e.g.][]{Tacconi2010}. At
$z>1$, and certainly around the epoch of peak cosmic star
formation ($z\sim2$), a significant number of disk galaxies are
clumpy, gravitationally unstable systems \citep[see
e.g.][]{Keres2009,Dekel2009b,Devriendt2010,Ceverino2010}, in which
star formation could be efficiently triggered by minor mergers
\citep[e.g.][]{Birnboim2003,Keres2005,Dekel2009b}. Since disks
form before bulges \citep[e.g.][]{Hatton2003}, the lack of the
stabilizing influence of a bulge may make these early disks even
more susceptible to minor-merger-triggered star formation
\citep[e.g.][]{Mihos1994,Hernquist1995} compared to their
modern-day counterparts.

Given the significance of this process in driving star formation,
black-hole growth and size evolution in massive galaxies, studying
the role of minor mergers across a large range in redshift is
essential. The most detailed insights are likely to come from
high-resolution spatially-resolved studies of local minor-merger
remnants using the HST, that employ sensitive tracers of star
formation like the UV. As noted in the introduction above, local
star-forming ETGs are excellent initial laboratories for such
spatially-resolved analyses that probe the star-formation laws
associated with satellite accretion (Kaviraj et al., in prep). At
intermediate redshift ($z<2$), the population of minor-merger
remnants can be studied using the deep, high-resolution imaging
offered by the HST \citep[e.g.][]{Kaviraj2011}, in legacy fields
such as CANDELS \citep[e.g.][]{Grogin2011,Koekemoer2011}. At
earlier epochs this exercise will likely require deep,
high-resolution imaging from future instruments like LSST and the
Extremely Large Telescopes. In any case, without a robust
determination of minor-merger-driven stellar mass and black hole
growth over cosmic time, our understanding of galaxy evolution is
likely to remain incomplete.


\section*{Acknowledgements}
I thank Martin Hardcastle, Roger Davies, Chris Conselice, Jennifer
Lotz, Meg Urry, Martin Bureau, Sara Ellison, Kevin Schawinski and
Sukyoung Yi for interesting discussions. I acknowledge a Senior
Research Fellowship from Worcester College Oxford.


\nocite{Yi2005,Kaviraj2007c,Kaviraj2011,Schweizer1990,Schweizer1992,Crockett2011,Merritt2001,Baldwin1981,
Veilleux1987,Kauffmann2003,Kewley2006,Shabala2012,Trujillo2011,Bluck2012}

\bibliographystyle{mn2e}
\bibliography{references}

\end{document}